\documentclass{ws-procs9x6}
\def\delslash{\,\,{\raise.15ex\hbox{/}\mkern-10mu \partial}}
\newcommand{\lapprox}{\raisebox{-0.5ex}{$\
\stackrel{\textstyle<}{\textstyle\sim}\ $}}
\newcommand{\gapprox}{\raisebox{-0.5ex}{$\
\stackrel{\textstyle>}{\textstyle\sim}\ $}}

\begin{document}

\title{A BCS Condensate in NJL$_{3+1}$?}

\author{Simon HANDS\footnote{supported by a \uppercase{PPARC} 
\uppercase{S}enior \uppercase{R}esearch \uppercase{F}ellowship}
 ~and David N. WALTERS}

\address{Department of Physics, 
University of Wales Swansea, \\ 
Singleton Park, Swansea SA2 8PP, U.K.}


\maketitle

\abstracts{
We present results from a lattice Monte Carlo study of the Nambu --
Jona-Lasinio model in 3+1 dimensions with a baryon chemical potential
$\mu\not=0$. As $\mu$ is increased there is a transition 
from a chirally-broken phase 
to relativistic quark matter, in which baryon number symmetry
appears spontaneously broken by a diquark condensate 
at the Fermi surface, implying a superfluid ground state. Finite volume 
corrections to this relativistic BCS scenario, however,
are anomalously large.
}

The proposal that at high baryon density a 
condensation of diquark pairs in degenerate
quark matter takes place, leading to color
superconducting ground states, has generated a 
profusion of theoretical and phenomenological interest~\cite{RW}. 
Most approaches 
assume the existence of a Fermi surface and then consider its instability
with respect to diquark condensation arising from an attractive $qq$ 
interaction. The resulting excitation energy gap $\Delta$ can be calculated 
self-consistently in close analogy to the BCS gap in electronic superconductors.
Model calculations yield gaps as large as  50 - 100MeV~\cite{BR}.

A first-principles calculation of the ground state of nuclear or quark matter
using the methods of lattice QCD, however,
is hampered by the ``sign problem'', ie.
$\exp(-S/\hbar)$ is no longer positive definite, and cannot be used
as a probability measure in importance
sampling the Euclidean path integral, 
once baryon chemical potential $\mu\not=0$.
It is still possible, however, to study simpler models of strong interactions
with $\mu\not=0$ using lattice methods \cite{SJH}. 
The Nambu -- Jona-Lasino (NJL) model, which closely resembles
the original BCS model, is particularly 
interesting in the current context, because it is the only simulable model
which exhibits a Fermi surface for sufficiently large $\mu$. The Lagrangian 
density is
\begin{equation}
L=\bar\psi(\delslash+m_0+\mu\gamma_0)\psi
-{g^2\over2}\left[(\bar\psi\psi)^2-(\bar\psi\gamma_5\vec\tau\psi)^2\right].
\label{eq:L}
\end{equation}
Here $\psi$ and $\bar\psi$ are isospinor ``quark'' fields 
carrying baryon charge;
note that both $q\bar q$ and $qq$ interactions are attractive in the isoscalar
channel. 

For coupling strength $g^2$ sufficiently large the vacuum 
ground state has chiral SU(2)$_L\otimes$SU(2)$_R$ symmetry 
spontaneously broken 
to SU(2)$_{isospin}$ by generation of a chiral condensate and resulting
``constituent'' quark mass $\Sigma\simeq g^2\langle\bar\psi\psi\rangle$. At zero
temperature this state is stable as $\mu$ is increased until at 
$\mu_c\sim\Sigma$ chiral symmetry is restored, at which point the baryon
density $n_B=\langle\bar\psi\gamma_0\psi\rangle$ rises from zero~\cite{SPK}. For
$\mu>\mu_c$ the ground state is thus relativistic quark matter with 
Fermi momentum $k_F\sim\mu$. A natural question is whether
$qq$ attraction in this phase leads to a diquark condensate spontaneously
breaking the U(1)$_B$ symmetry of (\ref{eq:L}) leading (since in this case the
broken symmetry is global rather than local) to 
superfluidity.

The NJL model in 2+1 dimensions 
has been studied extensively using Monte Carlo
simulations. While there is enhanced $qq$ pairing in the scalar
isoscalar channel for $\mu>\mu_c$~\cite{HM}, simulations with an explicit
 U(1)$_B$-violating source term
$jqq$ 
found no evidence for diquark condensation in the $j\to0$ limit~\cite{HLM}.
Instead, the condensate scales approximately as $\langle qq(j)\rangle\propto
j^{1\over\delta}$, the exponent $\delta$ increasing with $\mu$. Moreover, 
the quasiparticle dispersion relation in the spin-${1\over2}$ channel
is consistent with a vanishing BCS gap $\Delta=0$.
The interpretation is that 
long-wavelength fluctuations of the
complex phase of the $qq$ wavefunction destroy the condensate, but that
long-range
phase coherence remains (ie. correlations are dominated by ``spin-waves'' and
remain algebraic functions
of spatial separation) just as in the low-$T$ phase of the 2$d$ XY
model. Superfluidity in two dimensions is realised in Kosterlitz-Thouless mode;
namely, persistent flow patterns can only be
disrupted by creating a vortex/anti-vortex excitation and then translating one
member of the pair around the universe, which in the low-$T$ phase entails
an energy cost increasing logarithmically with system size.

This rather exotic scenario, namely gapless thin film superfluidity, is
inherently non-perturbative and is not exposed by self-consistent approaches.
It thus justifies the application of numerical simulations, and begs the
question of whether similar surprises will emerge from a study of the
phenomenologically-motivated case of NJL$_{3+1}$~\cite{HW}.  
In 3+1$d$ the model is
non-renormalisable, and hence its physical predictions are sensitive to the 
details of the ultra-violet cutoff. We have matched our lattice model to 
low-energy QCD following the procedure of Klevansky~\cite{SPK}, working 
to leading order in an expansion in $1/N_c$, where $N_c$ is the 
number of quark ``colors'', and  calculating 
$\Sigma$, $m_\pi$ and $f_\pi$ in terms of bare parameters $g^2$ and $m_0$.
Since we use staggered lattice fermions $\chi,\bar\chi$, we have $N_c=4$, 
in which case reasonable low-energy phenomenology emerges for the choice
$a^2/g^2=0.565$, $m_0a=0.002$, with inverse lattice spacing $a^{-1}\simeq1$GeV.
\begin{figure}[ht]
\centerline{\epsfxsize=3.9in\epsfbox{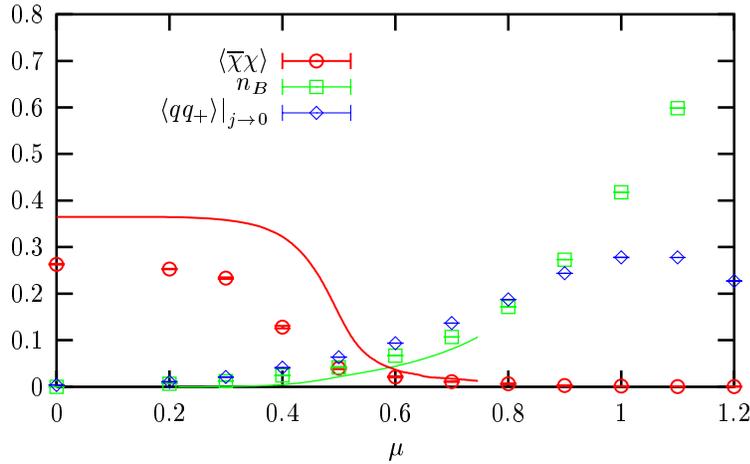}}
\caption{Chiral condensate $\langle\bar\chi\chi\rangle$, baryon density, 
and diquark condensate as functions
of $\mu$, showing both large-$N_c$ solution (solid curve) and simulation
results (points).
}
\label{fig:eos}
\end{figure}
Figure~\ref{fig:eos} shows results for $\langle\bar\chi\chi\rangle$
and $n_B$ and
confirms the transition between vacuum and quark matter at 
$\mu_c\sim\Sigma\sim{g^2\over2}\langle\bar\chi\chi\rangle$. 
The discrepancy 
between line and points is ascribed to $O(1/N_c)$ corrections and amounts to a
30\% effect. 
In contrast to
2+1$d$ where a sharp first-order transition is observed~\cite{SJH,HM}, 
the transition appears 
continuous in
the chiral limit $m_0\to0$, although this may be sensitive to the 
cutoff~\cite{SPK}.

To investigate the diquark sector, we introduce a source term
$j_\pm(q^{tr}q\pm\bar q\bar q^{tr})$ into the quark operator used for
measurements (since the simulation dynamics are performed with $j_\pm=0$, 
this is a
``partially quenched'' approximation) and define the following:
\begin{equation}
\langle qq_\pm\rangle={1\over V}{{\partial\ln Z}\over{\partial j_\pm}};\;\;
\chi_\pm={{\partial\langle qq_\pm\rangle}\over{\partial j_\pm}};\;\;
R(j_+)=\left\vert{\chi_+\over\chi_-}\right\vert_{j_-=0}.
\end{equation}
The nature of the ground state is then determined by
\begin{eqnarray}
\lim_{j_+\to0}\langle qq_+\rangle&=&0,\;\;\lim_{j_+\to0}R=1\;\;\;\;\;\;
\mbox{if U(1)$_B$ manifest}\nonumber\\
\lim_{j_+\to0}\langle qq_+\rangle&\not=&0,\;\;\lim_{j_+\to0}R=0\;\;\;\;\;\;
\mbox{if U(1)$_B$ broken}
\end{eqnarray}
where the vanishing of $R$ in the broken phase follows from the Goldstone
nature of the $qq_-$ boson in this case.

\begin{figure}[ht]
\begin{tabular}{ll}
{\includegraphics[scale=0.75]{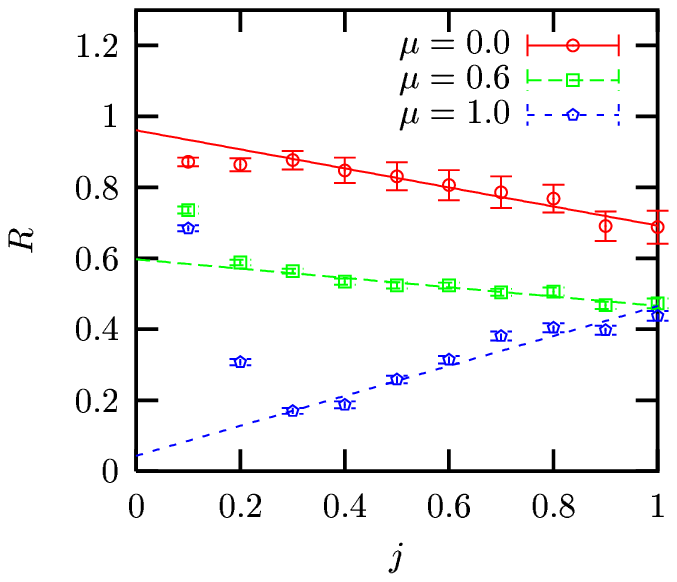}}&
{\includegraphics[scale=0.75]{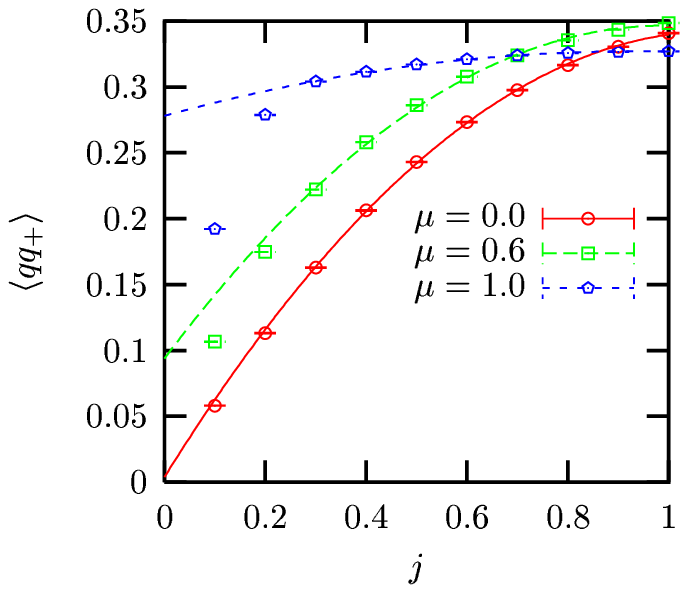}}
\end{tabular}
\caption{Susceptibility ratio $R$ (left) and diquark condensate 
$\langle qq_+\rangle$ (right) as functions of $j_+$ for various $\mu$.
}
\label{fig:results}
\end{figure}
Figure~\ref{fig:results} shows results for $R$ and $\langle qq_+\rangle$
from data taken on $12^4$, $16^4$ and $20^4$ lattices and then linearly
extrapolated to the limit $L_t^{-1}=T\to0$. For $\mu=0$ the data appear to
extrapolate smoothly to $R=1$, $\langle qq_+\rangle=0$ as $j_+\to0$. For
$\mu a=1.0$ by contrast, which lies in the region of high baryon density and
restored chiral symmetry, then if data with $j_+a\leq0.2$ are ignored the
extrapolated values are $R\simeq0$, $\langle qq_+\rangle a^3\simeq0.28$,
suggesting superfluidity. For intermediate $\mu$ the situation is not so
clear cut, but the results of a quadratic extrapolation of $\langle
qq_+\rangle$ to $j_+=0$ are shown in Figure~\ref{fig:eos}. For
$0.4\lapprox\mu a\lapprox0.9$ the condensate scales as $\mu^2$,
which is expected if diquark pairs within a distance $\Delta$ of a
Fermi surface of radius $\mu$ participate in the condensate leading 
to $\langle qq_+\rangle\sim\Delta\mu^2$. The value of $\langle qq_+\rangle$
at $\mu a=1.0$, together with an estimated $\Sigma\sim350$MeV, is thus
plausibly consistent with the gap values reported in [2].
The positive
curvature contrasts with the behaviour of the superfluid condensate in 
Two Color QCD in which $qq$ pairs are tightly bound and form a 
conventional Bose-Einstein condensate~\cite{KSTVZ}.

Since our interpretation of broken U(1)$_B$ symmetry at high baryon density
relies on discarding some data, the above conclusions are
necessarily provisional. As spontaneous symmetry breaking does not occur in
a finite volume, it is natural to attribute the depletion of $\langle
qq_+\rangle$ at small $j_+$ to the effects of a finite $L_s$. Note however
that the threshold $ja\sim0.3$ required for behaviour characteristic of the
thermodynamic limit seems anomalously large compared with that required 
to observe chiral symmetry breaking ($m_0a\sim0.002$) on the same volume, 
or indeed
for thin film superfluidity ($ja\sim0.1$) in NJL$_{2+1}$~\cite{HLM}.
We suspect that finite volume effects in systems exhibiting a Fermi surface are
unconventional, and indeed data on systems with varying $L_s, L_t$ suggest the
approach to the thermodynamic limit at small $j$ is non-monotonic~\cite{HW}.
Solution of the gap equation on finite systems shows that $\Delta(L_s)$
oscillates before approaching its infinite volume limit for $L_s\gapprox8$fm,
which translates to a $40^3$ spatial lattice for our model~\cite{ABMW}.

In future work we plan to study the quasiparticle spectrum which in principle
permits a direct estimate of the superfluid gap $\Delta$. It might also prove
interesting to choose different bare parameters, since a small change in the 
cutoff might actually change the chiral symmetry restoring 
transition to first order~\cite{SPK}, making it easier to distinguish the
properties of high- and low-$\mu$ phases. Finally, if our conclusion that 
NJL$_{3+1}$ is superfluid at high density perists, then it will be interesting
to examine the stability of the condensate as either temperature or
$\mu_{isospin}$ (which has the effect of separating $u$ and $d$ Fermi surfaces
and hence inhibiting pairing in the isoscalar channel) is raised from zero.

\end{document}